\documentclass[epjST]{svjour}
\usepackage{eurosym}
\usepackage{graphicx}
\usepackage{enumerate}
\usepackage{amsmath}

\begin{document}

\title{Multilayer adaptive networks in neuronal processing: Beyond neurotransmission networks}
\subtitle{}
\author{Adri\'{a}n Hern\'{a}ndez \and Jos\'{e} M. Amig\'{o} \thanks{\email{jm.amigo@umh.es}}}
\institute{Centro de Investigaci\'{o}n Operativa, Universidad Miguel Hern\'{a}ndez, Av.
de la Universidad s/n, 03202 Elche, Spain}

\abstract{
The connectome is a wiring diagram mapping all the neural connections in the brain. At the cellular level, it provides a map of the neurons and synapses within a part or all of the brain of an organism. In recent years, significant advances have been made in the study of the connectome via network science and graph theory. This analysis is fundamental to understand neurotransmission (fast synaptic transmission) networks. However, neurons use other forms of communication as neuromodulation that, instead of conveying excitation or inhibition, change neuronal and synaptic properties. This additional neuromodulatory layers condition and reconfigure the connectome. In this paper, we propose that multilayer adaptive networks, in which different synaptic and neurochemical layers interact, are the appropriate framework to explain neuronal processing. Then, we describe a simplified multilayer adaptive network model that accounts for these extra-layers of interaction and analyse the emergence of interesting computational capabilities.
} 
\maketitle

\section{Introduction}

\label{intro} In the last years considerable efforts are being dedicated to
provide insights into neural circuits in what has been called the connectome 
\cite{Sporns2005,Zador2012,Horn2013,Felleman1991}. The connectome is a map
of neural connections in the brain and may be thought of as its
\textquotedblleft wiring diagram\textquotedblright .

The connectome can be structural, if it describes anatomical connections
between parts of the brain or neurons, or functional, if it describes
statistical associations between activities in those parts. If we think of the structural
connectome as a road map, then the functional connectome corresponds to the
vehicles that travel the roads.

The knowledge of the neural elements and their neural connections can help
understand how the cognitive function emerges from the neuronal structure
and dynamics. This wiring diagram maps all the neural connections in the
brain and, at the cellular level, it provides a map of the neurons and
synapses in the brain.

The ideal framework to study and model the dynamics, topology and properties
of this type of synaptic connections is that of complex networks and
dynamical systems. Different approaches, from artificial neural networks to
biophysical models that take into account the biological reality
(conductances, response times, properties of synapses and dendrites, ...)
have been used to describe the dynamics of neuronal connections and
information processing in the brain; see \cite{Gerstner2014} for a
comprehensive description of neuronal dynamic models.

However, as some neuroscientists have pointed out \cite%
{Bargmann2013,Bargmann2012,Brezina2010,Marder2012}, knowledge and modelling
of the connectome, either structural or functional, are not enough to
understand how the brain processes the information, although they contribute
in a prominent way. There are other biological mechanisms, such as
neuromodulation, that reconfigure the connectome.

In neuromodulation \cite{Bucher2013}, a given neuron uses one or more
chemicals to regulate diverse populations of neurons, in contrast to
classical synaptic transmission, where one presynaptic neuron directly
influences a single postsynaptic neuron. Neuromodulators operate on several
time scales, and modulate and configure the connectome so as to determine
the processing of information. The connectome provides a minimal structure
and, on top of it, neuromodulators configure and specify the functional
circuits that give rise to behaviour.

Thus, neuromodulators add new computational and processing capabilities to
traditional synaptic transmission. First, they add extra-layers of
biochemicals that regulate or tune neuronal processing. Second, the temporal
scales of these extra-layers are different from classical ones. Third, these
extra-layers and classical synaptic networks interact in a complicated way.

We propose that the appropriate framework for modelling neuronal processing
is that of multilayer adaptive networks. Multilayer because in the brain
different synaptic and neuromodulatory networks interact to produce
behaviour, and adaptive because the topology of these networks changes
according to the dynamics.

The first goal of this paper is to highlight the limitations of the connectome
and neurotransmission networks to understand neuronal processing, as well as to
point out the need of using a new framework. We review examples in which the
same structural network can have different configurations and behaviours.

The second goal is to define the characteristics that a complex network
framework must have in order to provide a complete description of the
different neuronal information dynamics, scales and interactions that occur
in the brain. Multilayer adaptive networks, in which different synaptic and
chemical layers interact, are the appropriate framework to explain neuronal
processing and the emergence of interesting computational capabilities.

\section{Beyond neurotransmission networks}

\label{sec:2} The connectome is a wiring diagram mapping all the neural
connections in the brain. At the cellular level, it provides a map of the
neurons and synapses within a part or all of the brain of an organism. The
structural connectome provides the basis on which functional activity is
implemented and therefore shapes the functional connectivity.

On the way to unveil the connectome of the human brain, one of the ultimate
goals in neuroscience, some milestones have been achieved, e.g., the complete
connectome of the nematode C. elegans \cite{White1986,Varshney2011}, which comprises
302 neurons.

In recent years, significant advances have been also made in the study of
the connectome via network science and graph theory \cite%
{Achard2007,Alivisatos2012,Alexander2013,Baker2015,Barnett2009}. At the
cellular level, the nodes of the network are the neurons, and the edges
correspond to the synapses between those neurons. Therefore, graph theory is
the ideal framework to study the topology and dynamics of brain networks.
The combination of new tools to map and record neuronal patterns and the
computational techniques of network science has provided a new setting for
the study of the brain dynamics.

Many studies have been conducted to analyze the topology and dynamics of the
neuronal connectome. In \cite{Nicosia2013} the authors studied the growth
rules of the neuronal system of C.elegans. They found that the network
growth undergoes a transition from an accelerated to a constant increase in
the number of synaptic connections as a function of the number of neurons.
The transition between different growth regimes may be explained by a
dynamic economical model incorporating a trade-off between topology and cost.
In \cite{Towlson2013} graph theory is used to investigate the neuronal
connectome of C.elegans. The authors identified a small number of highly
connected neurons as a rich club interconnected with high efficiency and
high connection distance.

Clearly, connectome modelling and analysis play a salient role when it comes
to understand neurotransmission (fast synaptic transmission) networks.
However, neurons use other forms of communication as neuromodulation that,
instead of conveying excitation or inhibition, change neuronal and synaptic
properties.

As described in \cite{Nadim2014}, neuromodulators are released in modes
other than fast synaptic transmission and modify the neuronal circuit
outputs. They are the main factors in shaping behaviour by changing neuronal
excitability and synaptic dynamics and strength. The processes that are
subject to modulation include changes in probability of neurotransmitter
release, changes in transmitter receptors, modification of synaptic
strength, adding or subtracting ionic currents, and changes in voltage and
time dependence of channel gating. In doing so, they reconfigure the
connectome and provide adaptability of the brain functions.

Although traditionally neurochemical messengers have been classified as
small-molecule neurotransmitters, biogenic amines and neuropeptides, for the
purposes of this paper it is more appropriate to differentiate between
neurotransmitter (fast synaptic transmission) and modulator functions as in 
\cite{Brezina2010}.

In neuronal circuits, connectivity alone does not provide enough information
to predict circuits outputs \cite{Bargmann2012,Bargmann2013}. Neuronal
processing and behaviour are sensitive to intrinsic channels and electrical
properties that vary within and between cell types. Fast synaptic
transmission and biochemical processes interact to generate complex dynamics
in neurons and circuits.

Studies of neuronal circuits in invertebrate and vertebrate animals \cite%
{Bargmann2012} have revealed the ability of neuromodulators to reconfigure
information processing. They change the composition of neuronal circuits and
permit a circuit with a fixed number of neurons to produce many different
patterns of activity.

Then, the greatest challenge that we face to understand the brain is to have
new models that allow us to explain how the interaction of different layers
of neurotransmission, neuromodulators and genetic changes gives rise to
information processing. Moreover, without taking into account these
different interactions, it is impossible to explain many computational
functions observed in the brain.

\section{Multilayer adaptive networks in neuronal processing}

\label{sec:3} Network science has grown over the last decades to become a
relevant conceptual framework for the analysis of many real systems. A
tremendous progress has been made in the application of network models in
neuroscience. Modelling brain networks as graphs of nodes connected by edges
has provided major advances in understanding brain dynamics. From the
dynamic analysis of groups of neurons to the topological characterization of
large-scale human brain networks, network theory has become a fundamental
tool in neuroscience; see \cite{Froelich2016} and \cite{Fornito2016} for a
comprehensive description of network neuroscience.

Traditionally, the study of dynamical networks has covered either dynamics
on networks or dynamics of networks. In the first approach, nodes are
dynamical systems coupled through static links. This case includes dynamical
systems describing the dynamics in a phase space with no topological changes
between the nodes. In the second approach, network topology evolves
dynamically in time. This is the case of traditional complex networks, where
the focus has been put on analyzing the statistical properties that arise from
exogenous topological transformations.

In recent years, there has been a growing interest in adaptive networks,
i.e., networks whose links change adaptively with the states of the nodes in
an interplay between node states and network topology \cite%
{Sayama2013,Maslennikov2017,Wiedermann2015,Aoki2016,Aoki2015}. Two facts make
adaptive networks specially convenient for the study of natural
and social systems. First, dynamical processes on a network are sensitive to
the network topology, which influences the states of nodes. Second, the
states of the nodes feed back to the network topology creating a dynamical
feedback loop between topology and states of the network. In a neuronal
network, the firing rate of a neuron depends on the synaptic connections
(topology) and, in turn, the evolution of the synaptic connections and
weights depends on the neuronal activity. Furthermore, both processes
--neuronal activity and synaptic reconfiguration-- take place at different
timescales.

A typical adaptive, directed network with a fixed set of nodes is composed of
the following elements:

\begin{enumerate}[(i)]

\item A set of $N$ nodes $V=\{v_{1},v_{2},...,v_{N}\}$. Abusing notation,
node $v_{i}$ will be denoted by $i$.

\item Each node $i\in \{1,...,N\}$ has a state $s_{i}(t).$

\item The set of evolving links is encoded in a time-dependent, weighted
adjacency matrix $A(t)$ with entries $a_{ij}(t)$. In our case, self-links 
are excluded, so $a_{ii}(t)=0$.

\item Each link weight $a_{ij}(t)$ represents the relationship from node $i$
to node $j\neq i$ and is a function of $s_{i}(t)$ and $s_{j}(t)$. 

\item Node states $s_{j}(t)$ are a function of the sum of incoming weighted nodes,
that is, $\sum\nolimits_{i=1}^{N}a_{ij}(t)s_{i}(t)$.
\end{enumerate}

These systems change their states and topologies simultaneously according to
their interrelated dynamical rules. In a link removal $a_{ij}(t)\neq 0$
becomes $a_{ij}(t)=0$ while a new link is established when $a_{ij}(t)=0$
becomes $a_{ij}(t)\neq 0.$

In addition, until recently the majority of studies have focused on
single-layer networks, usually with a single type of node connected via a
single type of link. But in most biological systems multiple entities
interact with each other in complicated patterns that include multiple
layers of connectivity. Consequently, it became necessary to generalize
network theory by developing a new setting to study multilayer systems in a
comprehensive fashion \cite%
{Kivela2014,DeDomenico2016,DeDomenico2017,Menichetti2014}. Then, multilayer
networks are the suitable framework to study different networks that
interact to produce complex activities.

As we have seen, single-layer networks are represented using adjacency
matrices which, in the case considered, represent directed and weighted
relationships between the nodes of a network. Instead, multilayer networks
represent multiple dimensions of connectivity that, in the case of neurons,
can stand for different types of communication channels (neurotransmitters and
neuromodulators). A typical multilayer network has the following ingredients:

\begin{enumerate}[(i)]

\item A number $N$ of nodes (denoted by Latin letters $i$, $j$, ...) and a
number $L$ of layers (denoted by Greek letters $\alpha $, $\beta $, ...).

\item Node $i\in \left\{ 1,2,...,N\right\} $ in layer $\alpha \in \left\{
1,2,...,L\right\} $ has a state $s_{i\alpha }(t).$

\item A 4th-order, time-dependent adjacency tensor $M(t)$ with components $%
m_{i\alpha }^{j\beta }(t)$ which are the weights of the link from any node $i$ in
layer $\alpha $ to any node $j$ in layer $\beta $ in the network.
\end{enumerate}

\begin{figure}[t]
\centering
\includegraphics[scale=0.45]{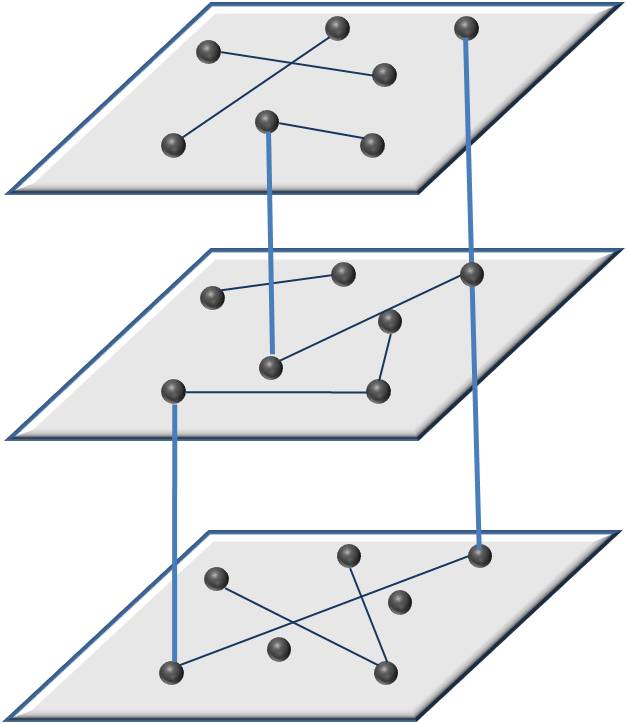} 
\caption{A multilayer network with intra-layer edges and inter-layer edges
that connect entities with their replicas in other layers.}
\label{fig:1}
\end{figure}

\begin{figure}[t]
\centering
\includegraphics[scale=0.45]{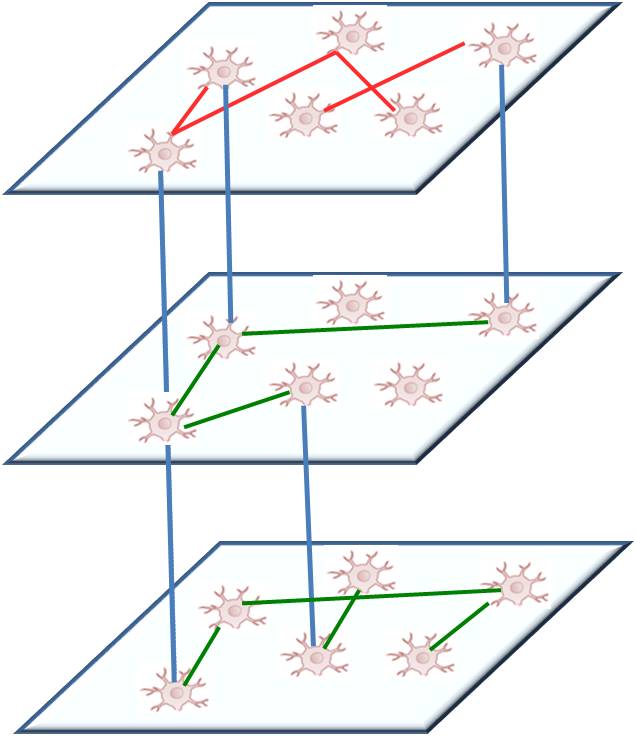} 
\caption{A multilayer network with one synaptic layer (red links) and two neuromodulatory layers (green links)
representing alternative modes of interaction between neurons, along with
the corresponding communication links between layers.}
\label{fig:2}
\end{figure}

In multilayer networks (see Fig. \ref{fig:1}) nodes can be connected by
different types of interactions: Intra-layer links connecting nodes within
the same layer, inter-layer links between the same nodes in different
layers, as well as inter-layer links between different nodes in different layers.
Multiplex networks are a special class of multilayer networks such that $%
m_{i\alpha }^{j\beta }=0$ if $\alpha \neq \beta $ and $i\neq j$, i.e.,
different layers are not interconnected except from each node to itself.

An interesting example of a multilayer network can be defined by extending
the connectome with synaptic and neuromodulatory layers representing
alternative modes of interaction between neurons, along with the
corresponding communication links between layers (see Fig. \ref{fig:2}).

The combination of multilayer and adaptive networks describes networks with
different interactions between their nodes, together with a dynamical
feedback between network topology and node states.

As we pointed out before, the connectome and single synaptic networks are
not enough to understand neuronal information processing. We summarize next
the characteristics that make multilayer adaptive networks the right
framework:

\begin{enumerate}[(i)]

\item In addition to fast synaptic transmission, neurons use other forms of
communication such as neuromodulation that change neuronal and synaptic
properties.

\item The extra layers of a multilayer approach regulate or tune neuronal
processing and operate on temporal and spatial scales different from the
fast synaptic ones.

\item The neuromodulatory layers interact with the fast synaptic
transmission layer in a complicated way, changing neuronal excitability
and synapses dynamics and strength.

\item The fast neurotransmission layer topology changes according to
neuronal and neuromodulatory activity.
\end{enumerate}

\bigskip

\section{Modeling neurotransmission and neuromodulation}

\label{sec:4} The connectome represents a network of potential neurons and connections
(synapses). Function, context and neuromodulatory networks shape and
reconfigure the connectome network allowing different paths of information
flow.

Neurotransmission is a wiring transmission that targets designated synapses
and produces localized responses. On the other hand, neuromodulation is a volume transmission
that diffuses through large areas of the neural tissue and affects multiple neurons. As a result, neuromodulatory networks are more complex than neurotransmission ones because neuromodulators act non-locally.

In \cite{Kopell2014} the authors hold that further understanding of brain
function and dysfunction will require an integrated framework that links
brain connectivity with brain dynamics. This expanded description is called
\textquotedblleft dynome\textquotedblright\ and includes the functional
connectome but expands the notion to the mechanisms involved in producing
and processing brain signals. Detailed biophysical models of neural activity
embedded in an anatomical network may be essential to examine the effects of
biological dynamics on functional connectivity.

As described in \cite{Daur2016} for the crustacean stomatogastric nervous
system, different regulatory mechanisms (synaptic and intrinsic neuronal
properties, neuromodulation and gene expression regulation) influence each
other to produce stable neuronal circuits.

The C.elegans connectome is considered in \cite{Bentley2016} as a multiplex network, with each 
node representing a neuron and with different layers
of connection (synaptic and neuromodulatory). The authors found
highly significant multilink motifs of interaction between the extrasynaptic
and synaptic connectomes, detecting locations in the network where the monoamines and neuropeptides modulate synaptic activity.

A simplified multilayer adaptive network model that accounts for these
extra-layers of interaction can be represented as follows:

\begin{enumerate}[(i)]

\item A number $N$ of neurons and a number $L$ of layers. Each neuron is
replicated in the rest of the layers, but with a different associated
dynamics. An adjacency tensor $M(t)$ with components $m_{i\alpha }^{j\beta
}(t)$ encodes the relationships from any neuron $i$ in layer $\alpha $ to
any neuron $j$ in layer $\beta .$

\item The first layer is the neurotransmission layer. Each neuron in this
layer has a state $s_{i1}(t).$ An evolving set of synapse weights is
represented by $m_{i1}^{j1}(t).$

\item The remaining layers are neuromodulatory layers. Each neuron in one of
this layer has a state $s_{i\alpha }(t)$ with $\alpha \in \left\{
2,...,L\right\} .$ An evolving set of neuromodulatory link weights is
represented by $m_{i\alpha }^{j\alpha }(t)$, $\alpha \in \left\{
2,...,L\right\} $.

\item The states in the first layer $s_{j1}(t)$ are a function of both the
sum of incoming synapses $\sum\nolimits_{i=1}^{N}m_{i1}^{j1}(t)s_{i1}(t)$ and the sum
of incoming interactions from neuromodulatory nodes $\sum_{\alpha
=2}^{L}m_{j\alpha }^{j1}(t)s_{j\alpha }(t).$

\item The synapse weights in the first layer $m_{i1}^{j1}(t)$ are a function
of both $s_{i1}(t)$ and neuromodulatory link weights in the remaining layers 
$\epsilon _{\alpha }m_{i\alpha }^{j\alpha }(t)$, $\alpha \in \left\{
2,...,L\right\} $, where $\epsilon _{\alpha }$ is a coupling parameter
between neuromodulatory and neurotransmission links.

\item The states in the remaining layers $s_{j\alpha }(t)$, $\alpha \in
\left\{ 2,...,L\right\} $, are a function of both the sum of incoming
neuromodulatory links $\sum\nolimits_{i=1}^{N}m_{i\alpha }^{j\alpha }(t)s_{i\alpha
}(t) $ and $m_{j1}^{j\alpha }(t)s_{j1}(t).$

\item The neuromodulatory link weights $m_{i\alpha }^{j\alpha }(t)$, $\alpha
\in \left\{ 2,...,L\right\} $, are a function of $s_{i\alpha }(t).$
\end{enumerate}

In this directed network, the first layer is a typical neurotransmission
layer with adaptive node states and synapses, where the link weights $m_{i1}^{j1}(t)$
represent classical fast-synaptic connections. The remaining layers contain
adaptive neuromodulatory states and links, where $m_{i\alpha }^{j\alpha }(t)$
are neuromodulatory link weights within layers. The interaction between
layers is of three types: Node states in the first layer are influenced by
the states of the same node in the neuromodulatory layers via $m_{j\alpha
}^{j1}$. Second, synapse weights in the first layer are influenced by the
same link weights in the neuromodulatory layers via $\epsilon _{\alpha }$.
Third, node states in the neuromodulatory layers are influenced by the state
of the same node in the first layer via $m_{j1}^{j\alpha }$. Adjacency
weights are different from $0$, $m_{i\alpha }^{j\beta }\neq 0$, when $\alpha
=\beta $ or when $\alpha \neq \beta $, $i=j$, and $\alpha =1$ or $\beta =1$.

Within the above setting, one can formulate detailed biological models
(e.g., Hodgkin-Huxley \cite{Hodgkin1952}) as well as more abstract models
(e.g., McCulloch-Pitts \cite{McCulloch1943}). For example, in an abstract
model the output of a neuron in the neurotransmission layer is given by the
weighted sum of its inputs. The neuromodulatory layers determine how
synaptic weights are updated. In such a model, neuromodulation can change
how a circuit function is achieved. The variables of a model are fixed
by the level of biological detail (concentration, membrane potential, state
of the neuron, etc.).

Regarding the dynamics, the evolution of a network is mostly described by differential equations (Hodgkin-Huxley, Fitzhugh-Nagumo, ...) or by difference equations (logistic maps, ...) depending on the kind and purpose of the model; see, e.g., \cite{Li2017,Maslennikov22017}.

As a final remark, by examining basic aspects of the multilayer adaptive models being
considered, several insights can be extracted. If neuromodulatory layers do
not depend on neurotransmission activity, then we have extrinsic
neuromodulation. In intrinsic neuromodulation, neuromodulatory layers are
not isolated from neurotransmission activity. It is also interesting to
compare the temporal scale of the neurotransmission layer with that of the
neuromodulatory layers. In most cases, neuromodulation is a slow process
compared to neurotransmission \cite{Fellous1998}.

In the next section we consider the importance of multilayer adaptive models
to analyze the computational capabilities of neuromodulators and to provide
a complete description of neuronal processing beyond the connectome.

\section{Computational capabilities of the multilayer connectome}

\label{sec:5}

As pointed out above, it is necessary to take into account the different layers of neuronal
communication to complete the connectome, both functional and structural. To this end, 
multilayer adaptive networks build the appropriate
framework to analyse how these layers interact to produce neuronal
processing. Next we focus on the computational capabilities added by these
interactions at the circuit level. The computational capabilities listed below can be materialized with two interacting layers, one for neurotransmission and the other one for neuromodulation. Other works have focused on aspects of
neuromodulators more related to behaviour \cite{Yu2013,Doya2008,Doya2002}.
Among those capabilities, we underline the following:

\begin{enumerate}[(i)]

\item The different time scales between neurotransmission and
neuromodulatory layers give rise to interesting phenomena. For example,
neuromodulatory activity in a slower process may fine-tune some properties
of the neurotransmission layer such as neuronal excitability and synapse
dynamics.

\item Because neuromodulation occurs in a diffusive manner, the same
neuromodulatory layer can tune different and isolated neuronal circuits
(neurotransmission layers).

\item Neuromodulatory layers can improve the functioning of neuronal
circuits. For example, in \cite{Holca2017} the authors found that the
neuromodulators trigger distinct changes in representations (through the
synaptic weights) that improve the networks classification performance. In 
\cite{Gutierrez2014} the authors demonstrate that neuromodulation of a
single target neuron may dramatically alter the performance of an entire
network or have almost no effect depending on the state of the network.

\item Neuromodulatory layers can ensure a reliable neuronal circuit function
despite changes in the parameters of the neurotransmission layer. These
extra layers regulate the correlation between parameters, providing
robustness to their variation \cite{OLeary2013,Marder2015}.

\item Neuromodulatory layers can reconfigure the neuronal circuit and even
change the function performed. This provides a very useful circuit
adaptability to the environment.

\item Neuromodulators possibly regulate the storage of new information in
neuronal networks. Neuromodulatory layers can maintain memory and activity
after reconfiguration of neuronal networks \cite{Reggia1999,Chaudhuri2016}.
\end{enumerate}

These facts show once more the limitations of the connectome to predict the output of
the circuit and the need to extend the connectome with additional layers of
neuromodulation that interact dynamically to reconfigure the connectome at
every moment.

This new approach will call for network theory and dynamical systems, along with
large computational resources, but it will be essential to understand the
complexity of the brain. Moreover, the study of neuronal circuits using
multilayer adaptive networks will provide clues and evidence of why the
processing of information in the brain is more complex and varied than the
one observed in artificial neural networks.

\section{Conclusions and future work}
\label{sec:6}

Important as is the study of structural and functional connectome networks, it is even more important to consider the additional neuromodulatory layers that condition and reconfigure the connectome. In line with the complex cognitive needs of the brain, there is no universal neural coding-decoding scheme but rather different layers of processes that add capabilities to information processing.

As we have seen in this paper, multilayer adaptive networks are an appropriate framework to study neuronal processing and to take into account all the communication processes that occur beyond neurotransmission (for example neuromodulation). Further work is still needed on concrete biological models of interesting phenomena via the multilayer approach.

Recent research has challenged the current premises in memory formation. The use of new techniques such as optogenetics has made it possible to differentiate between mechanisms of memory retrieval and memory storage \cite{Tonegawa2015,Ryan2015,Titley2017}. New theoretical and computational models, such as the one described here, will be required to explain these new observations.  

More generally, the use of multilayer adaptive networks for the study of neuronal circuits will lead to analyze the computational capabilities added by the additional layers. These computational capabilities (e.g., adaptability, regulation, robustness, degeneracy, memory, recurrency) will be key to understand the particularities of information processing in the brain and relate them to those of computers. 

Furthermore, cognitive scientists are calling for greater integration between neuroscience and artificial intelligence. This enlarged framework will allow to establish synergies and points in common between neuronal processing and current techniques such as machine learning. Machine learning needs new approaches to imitate how the brain learns and operates and, therefore, it may be relevant to consider how neuromodulation and biochemical communications condition the processing of information.

\begin{acknowledgement}
We thank very much our referees for their constructive criticism. This work was financially supported by the Spanish Ministry of Economy, Industry and Competitivity, grant MTM2016-74921-P (AEI/FEDER, EU).

\medskip

\noindent A.H wrote the draft. J.M.A. revised the draft. Both authors discussed the contents and agreed on the final version.
\end{acknowledgement}

\end{document}